\documentclass[sn-mathphys-num]{sn-jnl}

\usepackage{setspace}
\usepackage{graphicx}%
\usepackage{multirow}%
\usepackage{amsmath,amssymb,amsfonts}%
\usepackage{amsthm}%
\usepackage{mathrsfs}%
\usepackage[title]{appendix}%
\usepackage{xcolor}%
\usepackage{textcomp}%
\usepackage{manyfoot}%
\usepackage{booktabs}%
\usepackage{algorithm}%
\usepackage{algorithmicx}%
\usepackage{algpseudocode}%
\usepackage{listings}%

\theoremstyle{thmstyleone}%
\theoremstyle{thmstyletwo}%

\theoremstyle{thmstylethree}%

\begin{document}

\title[Stagnant Lakatosian Research Programmes]{Stagnant Lakatosian Research Programmes}

 \author*{\fnm{Johannes} \sur{Branahl}}\email{johannes.branahl@t-online.de}
\affil*{\orgdiv{Philosophisches Seminar}, \orgname{Universit\"at M\"unster}, \orgaddress{\street{Domplatz 23}, \city{48143 Münster},  \country{Germany}}}

 

\abstract{We extend Lakatos’s classical dichotomy of research programmes—progressive vs. degenerative—by introducing a neutral third category: the stagnant research programme. First, a critical examination of the primary literature with its often criticized definitional gaps justifies such a category. By deriving general criteria for stagnation, we clearly distinguish stagnant programmes from those that are progressive or degenerative. An empirical cross-check is subsequently employed for support: Both a series of examples from fundamental physics and a general analysis of today's research landscape also suggest on an empirical level the need to go beyond the traditional Lakatosian conception. Attributing stagnation is entirely in line with Lakatos' original intentions, which aimed not to hastily discard promising research but to exercise patience until the lifting of certain external constraints potentially enables empirical progress once again.}


\maketitle

\subsection*{Declarations.} 
\textbf{Availability of data and material:} not applicable \\
\textbf{Competing interests:} The author states that there is no conflict of interest. \\
\textbf{Funding:} The author did not receive support from any organization for the submitted work. \\ 
\textbf{Authors' contributions:} Single-author publication \\ 
\textbf{Acknowledgements:} We thank Ulrich Krohs und Florian Fabry for valuable discussions and all members of the colloquium for the discussion of publications of the Philosophisches Seminar (University of Münster) for a critical review.  
\section{Introduction}

Lakatos’s traditional dichotomy of research programmes—progressive vs. degenerative—suffers from two shortcomings. These binary categories appear to be neither mutually exclusive, nor collectively exhaustive. The first shortcoming became an early target of criticism given the concept's lack of definitional precision. According to Feyerabend, the definition of progressive programmes via "intermittently" empirical problem shifts represents an empty criterion allowing one to wait indefinitely: "why not wait a bit longer?" (1975, p.\,77). 
During long periods of absent experimental success, the boundaries between progress and degeneration seem to blur. This fact is closely intertwined with the second shortcoming, which has received little attention in the existing secondary literature on Lakatos and will be discussed in this article.

Facing the challenge of fully grasping the complexity and dynamics of today's research landscape through the Lakatosian categorization, one quickly notices that classifying programmes into black and white - into progressive and degenerative ones - reaches its limits in various cases, and the space of all programmes is not adequately and fully described by this binary division. We propose to expand the Lakatosian vocabulary by a third, neutral option - which we will call the stagnant programme - and to critically reflect on the structures of contemporary scientific practice. The question of a third category of stagnation encourages taking a nuanced perspective on the state and development of research programmes to precisely capture the space between progress and degeneration: The resulting extension does not treat stagnation as a deficiency but as a natural and non-condemnable phase in the research process of certain disciplines. As a protective mechanism, it bridges the gap between theoretical promises and the lack of empirical confirmation, without the theory being labeled as degenerative and hence, disregarded. Neutrality (the absence of empirical dynamics - stagnation) thereby implies that the programme is in principle not evaluable to be promising or discouraging. Four analytically derived criteria need to be fulfilled to demarcate stagnant programmes from all the others. As in the spirit of Lakatos, the diagnosis of stagnation can prevent a premature discard of promising research programmes and enables to bring the threefold categorization scheme more strongly into the present, rather than solely considering it as a historiographical tool.

We exemplify the category of stagnant programmes and motivate its necessity by concrete challenges of fundamental physics of this century, where the diagnosis of degeneration through missing empirical progress (and an enormous theoretical fruitfulness) may be inappropriate and misjudges the state of several research programmes: In principle, research could be conducted according to the best methodology, but progress is hindered not by internal misdemeanors, but by external, uncontrollable circumstances. In the end, the introduction of a neutral stagnation turns the previous categorization into both a mutually exclusive and collectively exhaustive framework for today's research, resolving the first shortcoming as a side effect, too.

 After critically examining the primary literature (Chapter 2) and analytically deriving four general criteria of stagnation in an analytical top-down-approach, where the ascription of degeneration is unwarranted (Chapter 3), support is drawn from a bottom-up empirical examination of today's research landscape. Particularly, current fundamental physics exhibits a high degree of alignment with the analytically derived criteria (Chapter 4.1). However, a bird's-eye view of the state of today's research in general reveals gaps in the classical Lakatosian dichotomy, too (Chapter 4.2).

\section{The Dichotomy of Progress and Degeneration}
\subsection{Foundations}

Let us recapitulate the Lakatosian methodology of scientific research programmes (MSRP) in a nutshell. Key concepts below include the programme’s hard core, its protective belt (or ring), and the positive and negative heuristics, which together determine whether it is progressive or degenerative.

The \textit{hard core} represents the non-negotiable set of basic theses and existing statements on which future research should be based. This foundation is not easily abandoned; doing so would be equivalent to abandoning the research programme itself. However, ongoing investigations typically bring data and phenomena that initially cannot be reconciled with the basic assumptions of the hard core. The inviolability of the hard core thus requires a flexible \textit{protective belt} that enriches the theoretical system with supporting hypotheses intended to reconcile emerging anomalies with the hard core. While the hard core enjoys immunity, the shape of the protective ring continually adapts to current threats to the hard core - modifications of the belt constitute the actual research process. The modus tollens of the research programme does not affect the hard core but is redirected to the protective belt. Together, the core and ring represent the evolving set of theories and explanations of the research programme over time. Models and theories thus exist not in isolation but are always interwoven into the structure of a research programme.

It is the task of \textit{positive heuristics} to provide general guidelines on how a complement or extension of the hard core should be carried out to explain the phenomena that have emerged over time coherently. Similarly, a heuristic in the negative sense can be formulated. This directive does not aim at gaining knowledge but rather excludes paths that would jeopardize the hard core. In short, these forms of heuristics are guided by the question how one should (not) proceed.

Research programmes are either progressive or degenerative, and a distinction is made between theoretical and empirical progress: Over the course of its development, the research programme forms a series of theories that, through constant adaptation in line with positive heuristics, successively surpass their predecessors in explanatory power and provide novel empirical findings (theoretically progressive problem shifts). If these predictions are also at least partially confirmed, there is an empirically progressive problem shift. The absence of such shifts indicates the degeneration of the programme.

The primary literature initially clarifies that a definitive attribution of degeneration is not readily possible because there can be no "instantaneous rationality" (Lakatos 1976, p.\,160), and an actual knowledge about the programme's state should exist only in hindsight. The high "methodological tolerance" (Lakatos 1970, p.\,71) exhibited by Lakatos' approach, which we will describe below, seemingly leaves the philosopher of science agnostic regarding the state of degeneration. The generality of the approach implies that no time scales for progress can be specified for the "intermittently progressive empirical shift" (Lakatos 1970, p.\,49) that characterizes true progress. Lakatos initially only conveys the central message that in every step of the research programme, there must be a theoretically progressive shift in problems when attempting to eliminate anomalies ("increase in content", ibid.). This not only serves the general applicability of the methodology of research programmes, but also reflects Lakatos' deep conviction, agreeing with Kuhn, that theories are born falsified (in his words being in an "ocean of anomalies," Lakatos 1970, p.\,48).  However, while Kuhn describes the absence of empirical progress by an irrational adherence to an outdated paradigm, Lakatos replaces it with a rational, momentarily justified adherence. The latter wants to prevent the demand for constant empirical progress from overly restricting the space for this adherence, even though the potential of a research programme may not have been fully exploited (Lakatos 1970, p.\,49). Despite the temporary lack of empirical success, a successful return of a research programme is conceivable, which ultimately outperforms other programmes in a sort of evolutionary competition. Conversely, this also means that not only a final degeneration but also no final victory of a programme over others can be attested without a considerable time gap in this competition. This relaxation of criteria is intended to ensure theoretical multipolarity and a fruitful competition for the best programme. This may even require the special protection of emerging research programmes (cf.\;Lakatos 1970, p.\,71 f.). However, lenient judgments need not lead to agnosticism. Lakatos opposes the claim that his approach represents "radical skepticism" (ibid.).

For even if real clarity about the state of research programmes predominates only ex-post, as outlined, Lakatos did not solely aim for a rational retracing of the history of science (as in Lakatos 1982, repeatedly claiming there was no "instant rationality"), but, in earlier days, also had an intention to bring the concept of progressive and degenerative research programmes into the present discourse. In other words: to anticipate problematic tendencies in contemporary research programmes (cf.\,Lakatos 1976, p.\,11). He aims to provide criteria for the elimination of entire research programmes, thus accepting an intervention in the course of the history of science through his considerations.
\\
By incorporating the category of stagnation, we will reintroduce a critical tool with normative power, which had been a central aspect of Lakatos' early works on the MSRP. As we will see, this extension becomes increasingly necessary in the age of modern research, where the complexity, scale, and resource demands of scientific inquiry often lead to programmes that appear stalled despite substantial theoretical development. The need for clear normative guidance in some disciplines becomes increasingly pressing, as researchers, funders, and policymakers navigate a landscape where empirical progress is not always immediate, and the continued viability of research programmes requires careful assessment beyond traditional categorizations. The extended MSRP discussed in this article aims to offer a framework for making these judgments.
\subsection{Quality and Dynamism of Research Programmes}
The call for action that can be derived from this text analysis can be specified as follows:  the claim of concrete criteria for the elimination of a programme fails because, on the one hand, unclear timeframes do not allow for an assessment. On the other hand, in the case of too long periods (whatever this means concretely for Lakatos), a hypothetical benign stagnation was indistinguishable from degeneration. Benign stagnation, however, is not equivalent to progress either. Circumstances are conceivable under which both attributions are inappropriate. These circumstances will be highlighted in the next section. 

Hence, we introduce the concept of a \textit{stagnant research programme} as a third state of a research programme. Interestingly, to the best of our knowledge, Lakatos uses the term \textit{stagnation} of the research programme only once, and without a deeper meaning (Lakatos 1976, p.\,11): "(...) it is stagnating if its theoretical growth lags behind its empirical growth, that is, as long as it gives only post-hoc explanations either of chance discoveries or of facts anticipated by, and discovered in, a rival programme (degenerating problem shift)." In this sentence, stagnation has got a pejorative character, is intended to grasp the lack of theoretical, not empirical progress, and is used synonymously with degeneration. However, the new category between progress and degeneration called stagnation should be understood quite differently: it expresses a neutral\footnote{Some remarks on the choice of "stagnation" as an appropriate term: The term is neutral in the sense that it does not imply negligence or a lack of effort on the part of the scientific community. Stagnation, as we define it, arises from inherent external or structural constraints despite the best efforts of scientists. While one may argue that “stagnation” carries a slightly pejorative undertone, partially inherited from economic doctrines, we intentionally chose it to reflect  the emotional realities faced by researchers working within such programmes, too. We thereby acknowledge the discouraging implications of working within a programme that has ceased to exhibit empirical progress.} evaluation of the current state of a research programme. More precisely, it is the absence of any evaluability on potential success or demise in a far future, caused by the absence of any empirical dynamics. Lakatos' classification requires that a research programme's dynamic developments move in one direction or another. Programmes that lack this dynamics cannot be classified by him. But equating standstill with degeneration shall be explicitly rejected - stagnation is not a reprehensible condition but precisely attests to the \textit{absence of degenerative elements}. Stagnation finally fills the white spots in the map of the research landsape, creating a mutually exclusive and collectively exhaustive categorization scheme.

 \begin{figure}[h!]
  \centering
  \includegraphics[width= 0.80\textwidth]{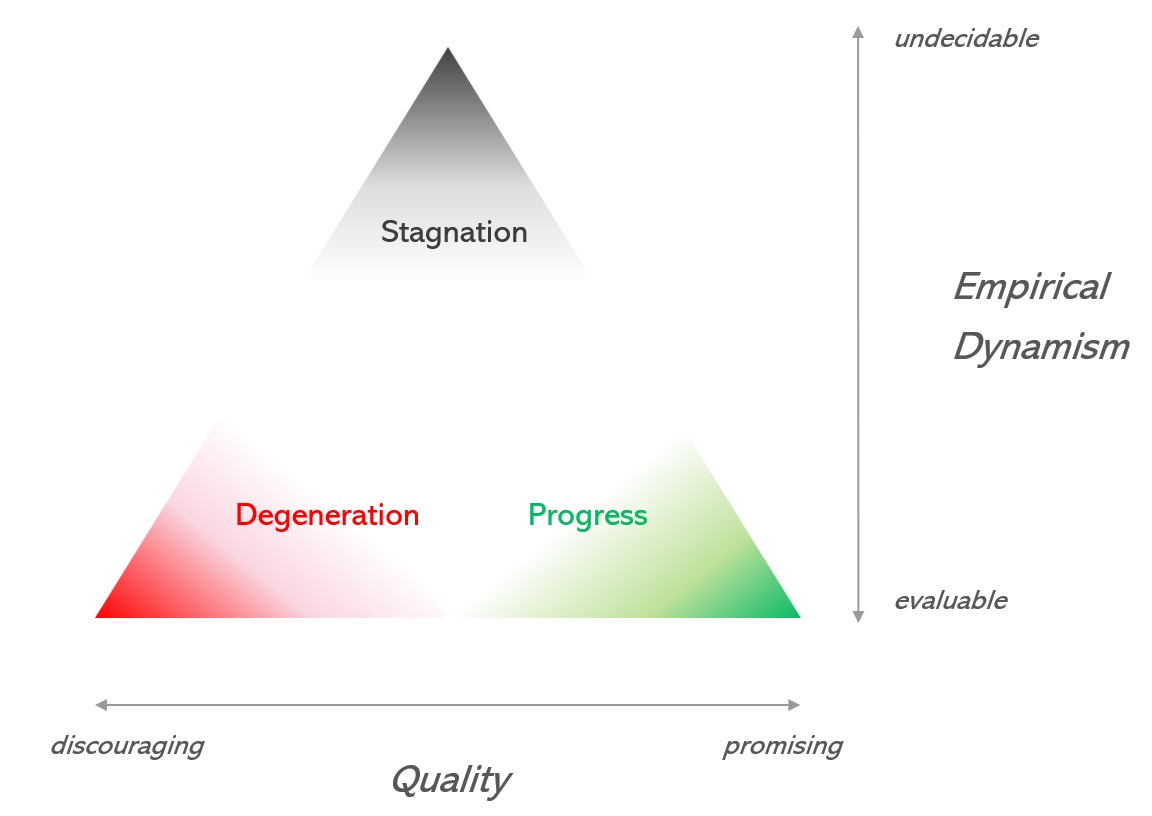} 
    \caption{Two-dimensional state space of research programmes - sorted by quality and empirical dynamism.  Programmes classified as "stagnant" cannot, by definition, be placed within the traditional dichotomy of "progressive" vs. "degenerative" because, due to their lack of dynamics, they cannot be evaluated at all. Instead of being an intermediate category, diluting the explanatory power of the Lakatosian categorization designed to be binary for simplicity and clarity, stagnation spans a second dimension in the state space being binary, too.  \label{ss}}
\end{figure}

To better illustrate the definitional sharpness of the concept of stagnant research programmes, Fig.\,1 shows a state space of programmes, where the introduction of stagnation adds a second dimension. In one dimension, the \textit{quality} of the programme is measured—classically dichotomous as "progressive" and "degenerative." However, to fully classify all contemporary programmes, a second dimension is needed, which measures (empirical) \textit{dynamism} and is also dichotomous: "dynamic" vs. "stagnant." This new dimension implies that stagnation is not seen as a vague intermediate category between progress and degeneration (both presupposing dynamics) and does not get relegated to redundancy\footnote{This sharp demarcation also ensures that the concept of stagnation (by being more than simply a delayed degeneration) can be abused by scientists to keep a degenerative research programme alive. The second binarity excludes any overlaps with what Lakatos might already consider a phase within a degenerative programme.}. Instead, it is recognized as an independent and clearly defined category, serving as the antonym to the dynamic developments in either a positive or negative direction. The triangular shape suggests that the quality of a research programme is somehow dependent on its dynamics. The greater the empirical dynamism of a programme, the broader the range of evaluable quality it can achieve (we remark that these clear geometric proportions are just an indicative scheme supporting the general idea. The concrete relationship between dynamism and evaluability may vary from programme to programme.). That is, they are evaluable because they are moving in a discernible direction, either toward improvement or deterioration. The base of the triangle represents programmes that are the most evaluable. These programmes exhibit high levels of experimental dynamics and can therefore be clearly classified into either quality category.  The tip of the triangle, representing the region of the state space where stagnant programmes are located, is where experimental progress vanishes entirely, making evaluability indistinct or impossible.

We finally remark a crucial asymmetry implying the unidirectional applicability of stagnation. The category of stagnation was introduced for scenarios of "pure theorizing" without experimental means of resolution. This is the only scenario of actual stagnation—in contrast to the opposite scenario of lacking theoretical explanations, appearing at first sight as stagnation, too: an overabundance of unexplained data without a suitable theory for a long timeframe causing a feeling of standstill and frustration. The difference lies in decidability—the possibility to evaluate the programme. When faced with unexplained experimental data, the problem lies with the failing theory. There is, therefore, an option of making a decision (remember that lacking evaluability is on the opposite a stagnant programme by definition): the theory must either be adjusted or discarded, and no neutral state of the research programme is imaginable. As there is never a severe hindrance to formulate new theories from the (newly protected) hard core, the absence of dynamics in a research programme can only be caused by empirical challenges. For what is "stagnation in theory development" but a guise for a series of fruitless premature ideas, discarded for their lack of empirical adequacy in the first place, before they could mature into a coherent theory? Such ideas are doomed to degenerate from the beginning, imposing decidability on the scientist from the outset. Thus, even phases of stagnation are only dictated by stagnating empiricism. The same asymmetry in phases of dynamics is merely another expression of what Lakatos recognized regarding the limitation to \textit{intermittently} empirical problem shifts.
But what makes, more precisely, a research programme stagnant?
\\
 Before clarifying this state with concrete criteria in Chapter 3, we dedicate a subsection to another prominent adversary of Lakatos and elaborate, why it is worth following the MSRP, although it never became \textit{the} evaluation scheme in philosophy of science. 

\subsection{Why Revive the MSRP?}

In contemporary philosophy of science, Lakatos’ methodology of scientific research programmes is often perceived as less relevant, especially when compared to Kuhn’s theory. It emphasizes social dynamics and epistemic values and may have been interpreted as more intuitive and more aligned with developments in the sociology and history of science. Additionally, Feyerabend’s more radical critique of scientific methods has overshadowed Lakatos in certain circles. Nonetheless, the MSRP—especially in an extended form that incorporates stagnation as a third category—possesses both significant descriptive and normative strengths, making it a valuable tool for our analysis.  While a detailed defense of Lakatos and a comprehensive comparison with Kuhn would warrant a separate study, we acknowledge that the MSRP never achieved the status of orthodoxy, partly due to its complexity and the rise of alternative frameworks.

The late Kuhn (1981), particularly as elaborated in response to Lakatos’ critique that his theory choice amounted to mere “mob psychology,” provides limited assistance in analyzing stagnation. Overall, while the qualitative interplay of epistemic values may indeed justify patience, as we propose for stagnant research, it lacks a structured mechanism for systematically categorizing the trajectory of research. Moreover, Kuhn’s five criteria for theory choice are useful in the coexistence of several productive theories to be compared, but reach their limits in the context of stagnant ones, being often singular. Making use of criteria such as simplicity, scope, consistency, and fruitfulness was purely non-empirical theory assessment in such cases and would require a sophisticated separate treatment to be applicable to static programmes. Of course, these values could help justify the continuation of a research programme during periods of stagnation. By prioritizing non-empirical virtues, an extended Kuhnian framework can rationalize the persistence of theories that are theoretically promising but empirically stagnant, naturally incorporating stagnation as a state of indecision or equilibrium. But ultimately, stagnation exceeds Kuhn’s framework far more than the Lakatosian one and does require much more effort than simply adding a third state or, in other words, a second coordinate in the state space of research programmes. The final criterion of accuracy, which assesses a theory’s closeness to empirical data, is effectively empty in the case of stagnation, as the very definition of stagnation involves the absence of empirical progress. 

On the other hand, the Lakatosian framework allows for a more natural incorporation of stagnation: A core aspect of the MSRP is the competition between research programmes, which Lakatos considers a fundamental driver of scientific dynamics. This competition allows for a distinction between progressive and degenerative trajectories, enabling an evaluation of the long-term development of scientific theories. By introducing a third category—the stagnant programme—this methodology is organically extended, filling a gap in the existing framework. Stagnation, understood as the absence of empirical dynamics while retaining theoretical potential, fits seamlessly into Lakatos’ system. This enhancement not only makes the MSRP more descriptively precise—creating a complete set of mutually exclusive categories simply by adding the opposite of a vivid competition—but also restores its original role as a practical guide for scientists. In earlier days, Lakatos’ aim was not merely to offer an ex-post rational reconstruction of the history of science but to provide ex-ante normative criteria for dealing with research programmes. This extension reaffirms that purpose.
\\ \\
While both approaches—Lakatos and Kuhn—remain interpretative, context-dependent, and particularly useful for textbook examples, we find the relatively simple extension of Lakatos’ MSRP to be a meaningful refinement. By distinguishing stagnant programmes, not only do the trajectories of all research programmes become reconstructible ex-post, but the ongoing or absent competition in the present can also be normatively assessed. This strengthens the original intent of the MSRP and makes it a valuable tool for analyzing modern scientific challenges. Reviving Lakatos’ MSRP seems worthwhile precisely because it has remained relatively underdeveloped. Unlike Kuhn’s ideas, which have been extensively explored, expanded, and critiqued, Lakatos’ framework has largely been left in its original form, creating a wealth of untapped potential for refinement and application. This lack of extensive elaboration means there is still significant room to adapt and expand his ideas to address contemporary scientific and philosophical challenges. In the current phase, which is dominated by alternative methodologies, revisiting and expanding on Lakatos's work offers an opportunity to give new impetus to the philosophy of science—while at the same time protecting the hard core of Lakatos's own programme in philosophy of science. Lakatos himself (1974) had already outlined such a recursive application of the MSRP in his Methodology of Historiographical Research Programmes.

\section{Analytical Top-Down Deduction of Criteria for Stagnation}

We begin by abstracting away from any particular programme that motivated this critique. Instead, this section includes general considerations based on the previous discussion of the primary literature of Lakatos, formulating criteria that a stagnant research programme has to meet. For this purpose, several demarcations are formulated, filtering the programme in a top-down approach such that the subset of stagnant research programmes remains in the desired sense of the word presented in this article. We briefly summarize the four criteria for stagnation first and continue with a more detailed elaboration.
\begin{enumerate}
\item[1)] Demarcation from Empirical Progress (Criterion of Pure Theorizing): There is a point in time beyond which the representatives of the research programme no longer consider the term "intermittently empirical problem shift" to be adequate (numerical values can vary greatly depending on the discipline) and the lack of empirical progress is identified as a problem.
\item[2)] Demarcation from Relative Degeneration (Criterion of Singularity): There is currently no alternative to the research programme fulfilling Criterion 1; there is no competition from up-and-coming research programmes and thus no dynamics allowing for an evaluation.
\item[3)] Demarcation from Absolute Degeneration and Closure (Criterion of Benign Epistemic Limits): Inherent and uncontrollable ("benign") epistemic limits prevent empirical progress in the research programme fulfulling Criteria 1-2 (in contrast to "malign" epistemic limits such as wrong predictions of a bad theory and methodological errors or, otherwise, the complete exploration of a research programme).
\item[4)] Demarcation from Insignificance (Criterion of Scientific Value):  Theoretical progress in the affected research programme that fulfils Criteria 1-3 is too significant to be abandoned (still strong social support, extended financial support, many active researchers of a new generation, fundamental importance for a discipline, ...).
\end{enumerate} 
\subsection{Elaboration on Criteria for Stagnation}
Let us dive deeper into the single criteria. First, the development of a third option between (empirical) progress and degeneration apparently requires the following separation:\\ \\ 
\textbf{1) Demarcation from Empirical Progress (Criterion of Pure Theorizing).} In this demarcation, being the most challenging one, the definitional problem of "intermittently empirical problem shifts" is aimed to be avoided. The arbitrary extension of the intermittency appears as an artificial adherence to a dualistic principle of rise and fall, the limits of which may not have been foreseeable at the time of its formulation. For the attribution of \textit{progress}, the permissible time until the empirical confirmation of the theory must be limited. The crucial factor in resolving the definitional vagueness of the dichotomous concept is only the existence of a period beyond which stagnation may occur. We emphasize that definitional clarity does not necessarily require concrete numerical values valid for any research programme. Whatever one postulates, Feyerabend wins. However, most scientific disciplines have an approximate, sometimes intuitive idea of a time frame - tailored to the specific requirements and environments of each discipline - within which empirical progress should take place before a state of degeneration is attested in the programme. 
Realizing these difficulties first allows a greater understanding of the heavily criticized definitional vagueness in Lakatos' original notion of a progressive programme, in his effort to maintain universality. 
 Nonetheless, the subjective awareness of degeneration of many researchers of such affected programmes is already sufficient to classify the arbitrary extension of "intermittently empirical problem shifting" as no longer appropriate. At the same time, the premature dismissal of the programme as a degenerative one - a direct consequence of the dichotomy - would not be adequate for several reasons given in the elaboration on Criteria 2-4. Criterion 1 thus ultimately corresponds to the intersubjective statement: "It is inappropriate to attribute a progressive state to the research programme given the lacking empirical dynamics so far". The consequence of lacking dynamics is then crystal-clear: in periods of (sometimes intersubjective diagnoses of) stagnation, the programme is not even evaluable by definition. While we aim to maintain universality, too, and thus refrain from specifying numerical time scales, we claim the existence of such time scales, tailored to the complexity of the problem under scrutiny, and hence circumvent the definitional gap in the Lakatosian dualism in a first step.
\\ \\ 
\textbf{2) Demarcation from Relative Degeneration (Criterion of Singularity).}  To distinguish from degeneration, it must first be ruled out that, unlike the new generation of the same research programme, representatives of another programme make progress towards empirical confirmation. If there were a dynamic competition between at least two programmes, there would inevitably exist a relationship between them that transforms stagnation into degeneration. However, if the competition has dwindled, degeneration may not be present: in this case, there have been no promising competing research programmes that could replace the existing stagnant one. Hence, in the majority opinion within the discipline, the current research programme would seem largely irreplaceable and without alternatives in its heuristics and hard core\footnote{Of course, the competition between individual elements of the protective belt on a theoretical level would not be a reason to end the stagnation of the entire programme.}. This condition may indicate that researchers are on the right track and follow the best scientific practices, but face hurdles of a different kind. 
 In special cases, Criterion 2 can be slightly relaxed. It is only necessary that the field of interest lacks disruptive alternatives that could decisively shift the research trajectory. There could be even a coexistence of several stagnant programmes, constrained by the same empirical challenges, but not undermining each other in their theoretical unfolding (validated in Chapter 4). Nonetheless, only one of them could be grounded on a sound hard core, whereas others would rely on misleading assumptions. Hence, the term of a singularity criterion is still valid.

In any case, this insulation in the landscape of research programmes could still represent degeneration if the programme is engaged in addressing ill-defined problems, where nothing needs to be explained. The programme would face an insurmountable hurdle, and the "problem" needs to be considered a degenerative element of the programme. These programmes must be sorted out in a third demarcation. 
\\ \\ 
\textbf{3) Demarcation from Absolute Degeneration and Closure (Criterion of Benign Epistemic Limits).} Three scenarios are conceivable in which different limits of an isolated programme are reached, but only the last one (3c) falls within the literal sense of stagnant research programmes:
\begin{enumerate}
\item[a)]  The end of the research programme is (assumed to be) reached, so in the absence of problems, there is not even theoretical progress. The tree of knowledge is harvested, and there is nothing new to discover, rendering the programme unassignable to any state. While this is not always immediately apparent, stagnation in the intended sense implies that the programme can in principle \textit{directly} contribute to an increase in knowledge by having unfulfilled goals.
 In case of 3a) we are not even dealing with a degenerative programme, the active programme is non-existent anymore after having reached a natural end. Instead, we are facing "closed programmes" in the sense of Heisenberg (1972) being not stagnant but finalized and foundational, thereby \textit{indirectly} contributing enduring value in their valid domains. They are not abandoned, but recontextualized. Heisenberg exemplified this category with classical Newtonian mechanics, the classical theory of heat, Maxwell's electrodynamics and non-relativistic quantum mechanics\footnote{The latter, controversial case, is discussed in Bokulich (2004) in greater detail. The author wants to thank one of the reviewers who brought the the concept of Heisenberg's closed programmes to his attention}.
\item[b)] The problem from which the research programme (or a part thereof) originated is unsolvable because it is ill-defined and does not pose a problem, but a pseudo-problem. In the case of the absence of relative degeneration (Criterion 2), absolute degeneration may be present: the programme's isolation (missing alternatives) does not necessarily imply a sound hard core. Incorrect basic assumptions and doubtful methodologies in theory development can thus generate pseudo-anomalies that cannot lead to stagnation\footnote{While labeling certain persisting issues as pseudoproblems risks appearing ahistorical (especially in the childhood of modern natural sciences) and pseudoproblems are often only recognizable retrospectively, once a superior alternative has emerged, this does not preclude their legitimate use as a standard of assessment in contemporary discussions. For example, the potential rise of pseudoproblems in foundational physics is not an arbitrary or retrospective imposition but a reflection of present concerns within the scientific community itself (see, e.g., Hossenfelder 2019), as we will detail out later. Addressing these concerns proactively, even in the absence of a definitive alternative, ensures a focus on research avenues with genuine potential for empirical or conceptual progress.}. We call it \textit{absolute degeneration} that does not require any competition with other programmes. 
\item[c)] \textit{Stagnation} occurs only in the case of limited experimental or cognitive capacities of representatives of the research programme, affecting the solution to correctly formulated problems. While the former is often easily determinable (limits of technical feasibility, economic constraints like financial possibilities), an intellectual limit of humans to these problems is speculative. Nevertheless, in both cases, the approach of the research programme is correct and should not be assigned to degeneration, but to stagnation\footnote{A remark with respect to the demarcation problem: pseudo-scientific hypotheses cannot be trapped in a stagnant research programme in the envisaged sense of the term, because the argument of benign epistemic limits can never be applied here. There will be always an absolute degeneration.}. In most scientific branches and in history of science in general, no one wishes and wished to pursue a research programme, if benign epistemic limits indicate that it would probably never make empirical progress, at least for several generations of researches, far beyond time scales Lakatos himself might have had in mind while developing his framework. Anyone working on a research programm used to do so in the expectation that progress eventually occurs and, if so, regarded it as progressive, Some regions of today's research landscape, as we will work out below, reveals a different situation. 
\end{enumerate}
Criterion 3 thus distinguishes final success (3a - completion) and internal failures (3b - absolute degeneration) of a research programme, which can in principle be prevented or cured, from external, uncontrollable forces - the intellectual nature of the human being and the society they live in (3c). These external forces shall be called benign epistemic limits, a strong indicator for stagnation, that even programmes with most promising theories and excellent methodology could face. Stagnation can be permanent due to these inherent limits, but it can also be transformed into progress in some very distant future since these may be limits of a scientific era and not fundamental epistemic limits. Benign epistemic limits hindering an "intermittently empirical problem shift" used to suggest, in the classical dichotomy, that one has to give up on solving problems facing these limits. This unconditional surrender would blight any future theoretical development potentially overcoming these benign epistemic limits, if not strictly inherent and not persistent forever. It encourages the researcher to keep going, although progress seems impossible. Thus, as Lakatos wishes for both the progressive and degenerative states, stagnation generally remains a provisional attribution that may solidify only through the historical retracing of the discipline. \\
After this filtering it is confirmed that in the stagnant research programme there is in principle still something to be discovered. However, there is still the option  that acceptance of the limited experimental reach and intellectual capabilities of Homo sapiens leads to resignation or inactivity or that the scientific value is simply to small to keep the programme alive (reminiscent of degeneration), necessitating a fourth filter:
\\ \\ 
\textbf{4) Demarcation from Insignificance (Criterion of Scientific Value)}. Lakatos writes about degeneration, among other aspects: "In the methodology of research programmes, the pragmatic sense of rejection (of the programme) becomes crystal clear: it is the decision not to work on it anymore" (1974, p.\,152, Fn 245)\footnote{This rejection is not caused because the state of Criterion 3a) is reached.}. A complete rejection of a programme as the ultimate consequence of stagnation presumably occurs rarely in its entirety in this specific subset of programmes filtered in all Criteria 1-3. Instead, a small grouping possibly remains, which, despite the majority's rejection of these activities, continues to dedicate itself to the programme. This form of stagnation still has degenerative characteristics; it resembles the gradual end of research programmes that lag behind in the competition for the greatest empirical adequacy, steadily losing followers, with the difference that in this case, even competitors are absent. However, in the case of the neutral attribution of stagnation, as envisaged in this paper, an opposite criterion is needed, according to which the research programme has already recovered from phases of doubt or resignation and is actively pursuing open questions intensively and with considerable effort. For the scientific discipline, solving the problems of the affected research programme is therefore too significant to bury it. In short, Criterion 4 asks: while there's still more to discover, is it meaningful, and do we want to invest significant effort into these potential findings that might be out of human reach forever? 
\subsection{Embedding of Criteria for Stagnation into Previous Debates}
We have seen that defining stagnant programmes first involves a relatively simple demarcation from progressive ones, requiring only the closure of the much-criticized definition gap of "intermittently" empirical problem shift. In contrast, the demarcation from degenerative programmes revealed the multitude of their variants - three criteria were needed to filter out those programmes whose characterization no longer carries pejorative connotations, and even their refinement will be needed when it comes to an empirical validation of Criteria 2-4. In other words, most progressive programmes are alike, each non-progressive programme is stationary in its own way.

Several other authors reflected about possible ways of stagnating research\footnote{The sources discussed in the following mostly do not explicitly mention the term of stagnant programmes, but elaborate on the same idea.} for longer periods within the entire twentieth century, confirming the previously derived set of four criteria: in the 1990s, the idea that a culmination point of many sciences could soon be reached gained broader attention (Horgan 1996). Horgan's work on the end of science can be translated into the terminology outlined in this chapter that an increasing number of scientists from various disciplines will find themselves in stagnant research programmes in the coming decades. Particularly concerning fundamental physics, he observes a departure from empiricism, as contemporary theories increasingly elude experimental accessibility (similar to Criterion 3). For this, he defines "ironic science" as closer to philosophy, literary criticism and even literature itself offering points of view and opinions that do not converge to the truth.

However, the debate about such an end to scientific progress is considerably older. It particularly supports the Criteria 3 and 4 of benign epistemic limits and scientific value and will be briefly retraced here. Bury (1932, p.\,1) early on noted that stagnation may not necessarily result from a complete exploration of a field: "How can we be sure that some day progress may not come to a dead pause, not because knowledge is exhausted, but because our resources for investigation are exhausted - because, for instance, scientific instruments have reached the limit of perfection (...)?" He also speculates about the scenario of intellectual limits of the human species (ibid.). Stent later adopted a similar stance, differentiating between various disciplines. He believed that biology might one day complete all empirical findings (supporting Criterion 3a), indicating no stagnation. In contrast, physics, with no inherent limits, could, in principle, explore things on ever higher energy and ever smaller length scales but faced principal limitations (supporting Criterion 3c) in empirical reach due to physical, cognitive, and economic constraints (Stent 1969, p.\,74). According to Stent, one should thus look for a neutral standstill, according to the criteria he previously derived, primarily in physics. In a later work, he justifies the consideration of imminent stagnation through the paradox of progress (Stent 1978): it is contradictory to conclude from the golden age of progress that this successful course continues indefinitely. The very limitation of things that humans can know and demonstrate would imply that the success story of science in the 20th century could come to a sudden end in the near future. 

In the same year, Rescher joined Stent's position, stating that the process of knowledge is at least infinite on an abstract level but societal and political acceptance limits progress due to increasing financial resources with decreasing epistemic returns. In light of the previously derived criteria, he bridges Criterion 3 to Criterion 4: ultimately, economic constraints, alongside technical feasibility, are the essential reasons for a fundamental limit to the experimentally achievable. Even if a research programme is highly significant (Criterion 4, therefore, remains unaffected by economic constraints) - beyond certain costs, stagnation is inevitable, despite the best scientific methodology (cf.\;Rescher 1978). Therefore, good scientific practice includes resource-efficient and economically efficient planning of experimental proof as a much more important operation mode.

Dawid (2019, p.\,105) addresses the time period of absent empirical confirmation or disclosure as follows, thereby even concretizing the timeframe of missing empirical evidence (Criterion 1) by more than half a century: "It may still make sense to ignore intermediate epistemic states between ignorance and conclusive knowledge in contexts where they last only for a brief period of time before the case is settled based on conclusive empirical evidence. In contemporary fundamental physics the typical time scale for that intermediate state has grown beyond the length of a scientific career." 

Although Criterion 2 was introduced rather for technical reasons to separate relative degeneration in the competition of programmes (where a stagnant programme is outperformed by a progressive one, inevitably causing degeneration by their relation to each other) from an isolated stagnation, this requirement also finds support on a completely different frontier: Dawid's (2013) no-alternative argument, with a very similar statement, is used to justify the complete abandonment of empirically progressive problem shifts as a (non-negotiable) condition for progress in some future. With this \textit{post-empirical} approach to theory evaluation, as it was named soon after by Huggett (2014), Dawid addresses especially string theory and other areas of fundamental physics whose experimental requirements permanently lie beyond economic and technical reach. This extensive departure from established standards will be initially set aside in this chapter. However, in cases of hopeless stagnation, this approach could represent the only way out of the pessimistic perspectives of the aforementioned authors, even if it initially appears as capitulation and reminiscent of Horgan's (1996) ironic science. 

We note that the derived criteria can be legitimized by referring to various analytical contributions to debates of the past hundred years. But we are not done at this stages: the top-down approach to deriving criteria for stagnation, while logical, may overlook the complexities of actual scientific practice. We continue with discussing how these criteria would interact with the messy realities of research, where programmes often display elements of both progress and degeneration simultaneously: Criteria 1-4 must prove themselves in practical application. This step will be undertaken in the following section.

\section{Empirical Cross-Check of Criteria for Stagnation}
In this section, the logic of characterising stagnant research programmes is reversed: the following empirical test of the introduced distinction between progress, stagnation and degeneration will initially be conducted using some specific examples. Additional suitable criteria will be identified during this process. Subsequently, a shift will be made to a holistic perspective that aims at comprehensively examining the tendency towards stagnation in contemporary science and justify the introduction of the concept of stagnation.

\subsection{Examples of Stagnant Research Programmes}

As already hinted in the introduction, the debate around stagnant research programmes is mainly motivated by fundamental physics and its state in the twenty-first century. Going outside of the realm of fundamental physics with particular use cases is, however, beyond the scope of this article, since complicating and discipline-specific features may occur.  

The following candidates for stagnant research programmes have remained without empirical successes to this day. They pertain to the problems of the two standard models in fundamental physics - particle physics and cosmology, based on quantum field theory and general relativity, respectively - along with their unification in quantum gravity\footnote{To avoid misunderstandings, we define physics as a fully empirical branch of science, where no kind of non-empirical theory evaluation suffices to confirm them and to generate an actual progressive problem shift in the Lakatosian sense.} 

Physics beyond the Standard Model of particle physics (BSM) addresses all the missing explanations in the Standard Model of the 12 fermions, 12 gauge bosons, and the Higgs boson. Theories within this research programme involve, among others, the generation of neutrino masses, the nature of dark matter (which exceeds known matter by a factor of five), a grand theory unifying the three interactions understood on quantum-level (GUT), the strong CP problem of quantum chromodynamics, and the hierarchy problem of the Higgs mass. The Standard Model itself enjoyed remarkable empirical success, culminating in the discovery of the Higgs boson in 2012 (Large Hadron Collider, Geneva) as the last missing piece. Since the 1970s, a three-digit number of BSM theories have been proposed, which all remain empirically unconfirmed. In SM physics, theory and experiment often went hand in hand - in BSM physics, they do apparently not. While there are certainly degenerating branches of BSM physics, as the multitude of experimentally excluded regions of the parameter space of supersymmetric extensions (mostly within reach of today's accelerators), we will focus on those hypothetical particles living in the so-called "desert" between contemporary colliders (roughly $10^4$ GeV) and the scale of a grand unification (GUT scale, roughly $10^{15}$ GeV)\footnote{This desert extends in the opposite direction, too. The hypothetical axion could only be compatible with the observed gravitational lensing in the elliptical galaxy HS 0810+2554 if they have a particle mass of $10^{-22}$ eV (Amruth et al. 2023). The axion as a solution for the strong CP problem, however, is heavily criticized and is an illustrative rather than plausible example for energies being unaccessible in principle.}. This widely unaccessible energy spectrum is the most convincing element of BSM physics to discuss the option of a stagnant research programme. The envisaged "Future Circular Collider" as a natural successor of the LHC, might reach $10^5$ GeV in some forty years. Alternative collider technologies like the plasma wakefield technology (Shiltsev 2020) could reach $10^7$ GeV in a very generous estimation. Thus, the problem of maybe centuries of lacking empirical evidence in the (direct) search for, e.g., leptoquarks with energies at the GUT scale is obvious\footnote{Harlander, Martinez, and Schiemann (2023) elaborate on the "end of the particle era" and only have some hope about indirect detections of new particles in the strongest sense of the word. }. And so is the frustration: the correct theory, perhaps up to details and numerical values of parameters, might have been already found - there might be nothing wrong with the methodology of the research programme and there might be no reason to leave it. Still, the intersubjective intuition that BSM physics is no progressive research programme anymore is omnipresent, from publications in refereed journals to international newspaper and books (Lindley 1993, Horgan 1996, Hossenfelder 2018). But why calling it degenerative?  \\
Distinguishing from degeneration in Criterion 2 must be approached with caution: the plurality of approaches in BSM physics and seems to create some tension with the singularity condition. However, the hard core of the programme—phenomenologically, the Standard Model and mathematically, the framework of renormalizable, perturbative quantum field theory —along with some conservation laws and the principle of naturalness\footnote{See Craig (2022) for an introduction.}, still appears today as the only option for progress in BSM physics. The particles and parameters of the Standard Model are outstandingly empirically confirmed (to some extent, the overwhelming success of the Standard Model is a considerable reason for the stagnation in BSM physics), and a more mathematically rigorous or even axiomatic version of quantum field theory in four dimensions is yet to emerge (turning it into a closed field of research in Heisenberg's sense, but also creating a different research programme with its individual hard core). Thus, despite the diversity of approaches, they can still be seen as belonging to the same overarching research programme when viewed through the lens of their shared hard core. The coexisting approaches (e.g., SUSY-based models or those without SUSY, with additional gauge groups or without, ...) represent different supplements in the protective belt rather than distinct research programmes. This perspective allows us to view the plurality of approaches as different attempts to address specific anomalies of the Standard Model within the same programme, rather than as violations of singularity.
\\
However, in distinguishing from absolute degeneration according to Criterion 3, challenges arise: assuming that all problems of the Standard Model are well-defined, both experimental and intellectual limits of humanity are in question. In particular, the high collision energies of accelerators or sensitivities of detectors that could detect the postulated particles, sometimes many orders of magnitude beyond the current standard, pose a central technical and financial hurdle for an empirically progressive problem shift.
 However, this assumption is not necessarily justified: first, it is unclear whether dark matter has got a particle character, or astronomical observations align better with modifications to the theory of gravity. Therefore, it is possible that the mystery of dark matter is more likely to be solved within the research programme of cosmology and gravity. Second, regarding the strong CP problem and the hierarchy problem, which represent anomalies to the principle of naturalness, it is debated whether both fundamentally pose a problem or are solvable in no research programme (Giudice 2017, Hossenfelder 2021). There is no guarantee for the validity of the so-called technical and 't Hooft-naturalness as metatheoretical principles in BSM physics. Perhaps the BSM programme needs to be liberated from degenerative elements first before being declared stagnant. This is why we concentrated on the benign epistemic limit of unreachable energies up to the GUT scale in the first place, restricting the option of stagnation only to some branches of the entire complex research programme. Criterion 4 is unquestionably fulfilled due to high financial expenditures (20 billion dollars envisaged for a "Future Circular Collider") and thousands of particle physicists worldwide, as well as the socially recognized importance of the Faustian aim to understand "what holds the world together in its inmost folds"\footnote{The concept of stagnant programmes could also influence the thorough evaluation of future "Big Science" projects like the Future Circular Collider (cf. Massimi 2021) by emphasizing the risks of investing in areas that currently show little or no dynamism. In the context of BSM physics, where experimental breakthroughs have been elusive for decades, classifying such research as stagnant suggests a cautious approach. Rather than committing billions of taxpayer's money to a project that may not yield results, evaluators might adopt a "watch and wait" strategy to delay large-scale investments until there are clear signs of renewed dynamism in the field indicating a productive and progressive future direction. This approach would prevent premature spending in potentially stagnant areas and ensure resources are allocated when there is a greater likelihood of meaningful progress.}.

Finally, we discuss what is often referred to be the Holy Grail of physics (referring already to Criterion 4), being the unification of general relativity and quantum theory. Its incommensurability was noted shortly after the formulation of quantum theory, and it most probably presents a benign, well-defined problem beyond the current and future experimental range. In the 1950s and 1960s, Feynman and DeWitt, among others, attempted to describe gravity quantum mechanically. In a subsequent epoch, loop quantum gravity became another approach developed by Ashtekar (1986), and Rovelli and Smolin (1987). It is based on the idea of breaking down space and time into smallest quantized units and uses a mathematical structure called loops. String theory, if ever considered empirically testable, provides an alternative description of gravity, but seems to face the most persistent experimental hurdles. Until 2024\footnote{See, e.g., (Oriti 2009) for further information on quantum gravity.}, significant theoretical progress at many frontiers was achieved. But so far, there is no experimental evidence for either theory of quantum gravity - there are not even any widely accepted ideas for experiments.  For example, it would take a particle accelerator of astronomical dimensions to reach the energy scales at which quantised gravity would make itself felt - even beyond the GUT scale at which BSM physics would have to operate. Quantum gravitational effects may have been at work during the extreme conditions in the first moments of the Big Bang. However, reaching this so-called Planck scale directly is likely to represent a fundamental, well-defined epistemic limit. Quantum gravitational effects in black holes are just as inaccessible. At the singularities in their interior, the physics we know collapses, but we cannot look inside any of them and investigate what happens there: a clear epistemic bondary. Even a measurable gravitational interaction between quantum-mechanically entangled objects requires masses that make artificially induced entanglement in the laboratory impossible.  
Given the recurring features of fundamental physics programmes that can be considered stagnant according to the analytically derived criteria, some additional criteria can be added, which may or may not be fulfilled in the case of stagnation:
\begin{itemize}
\item Regarding the Criterion 2 of singularity: if the programme is not lacking alternatives, stagnation still exists if the problem solution can be shifted to another stagnant research programme (e.g., dark matter). With respect to quantizing gravity, the relaxed version of Criterion 2 applies: theoretical progress in string theory does not undermine loop quantum gravity so far, and both programmes are facing the same benign epistemic limits of their falsifiability.
\item Regarding Criterion 3 of benign epistemic limits: the research programme is already highly developed and addresses ultimate, fundamental questions (e.g., quantum gravity as the \textit{Theory of Everything}, smallest building blocks of matter, geometry of the Universe) strongly challenging the cognitive limits of human beings.
\item  Regarding Criterion 4 of scientific value: the importance of the research programme is expressed through high societal relevance, the fundamental nature of its questions, and thus the regular approval of new funding.
\item To distinguish from absolute degeneration of the kind 3b): the research programme has been thoroughly examined for pseudo-problems as degenerative elements.
\end{itemize} 
The sceptical reader might argue that the presented examples of stagnant research programmes - BSM physics and quantum gravity - simply provide examples of progressive programmes with underlying theories being extremely hard to test. But what is the benefit of the diagnosis of stagnation for researchers?  Unlike other critics, we see our criticism of fundamental physics using the language of stagnant research programmes as therapeutic rather than disastrous. This category becomes particularly important as an opposition to more radical ideas, by asking for patience in times of uncertainty and frustration.  Both Hossenfelder (2018) and Dawid (2013) present ideas which, due to the lack of empirical findings, suggest a complete departure from previous experimental practice in accelerator physics and thus from the research programme of BSM physics as a whole (in the way it is carried out today). Hossenfelder's debate, which is important and fruitful in principle, concerns many justified points about possible pseudo-problems and the excessive importance of aesthetic criteria in the selection of new theories - weak points and potential degenerative elements of BSM physics. She criticizes the lack of alternative programmes (Criterion 2) and calls for greater mathematical rigor. Nevertheless, the largely metatheoretical argument ignores the plausible possibility that there are correct proposals within existing theories that escape scrutiny due to benign epistemic limits (Criterion 3). Without concrete predictions, she concludes, constructing larger accelerators is a waste of time. Leaving a programme has always been a sign for or a consequence of its degeneration, but Criteria 2-4 contradict this impression: someone who believes that a research programme has become degenerative due to empirically unconfirmed theories will leave that programme sooner or later, its reputation suffers. The course of action indicated when some believes a programme is stagnant, is the opposite: the diagnosis is triggered by the same absence of empirical progress, but criteria for stagnation may provide convincing reasons for it, and thus one decides to continue working on it, or at least pauses and restores its reputation. 

 The later academic debate about non-empirical theory assessment initiated by Dawid\footnote{We remark that Lindley (1993) brought Dawid's key insight of a need for non-empirical theory assessment into the broader public already two decades before the detailed elaboration on concrete criteria by claiming that contemporary physicists need to finalized their discipline in the way the ancient Greeks did in the childhood of physics: simply by rational, analytical thinking, returning to the ancient methodology, which is doomed to failure according to him. He calls it the "end of physics" - for Lindley a sign for the degeneration of an entire branch of science.} (see end of Chapter 3) initially recognizes empirical stagnation of promising theories that may persists forever as a challenge to the philosophy of science. However, criteria for the non-empirical testing of theories are very controversial and cannot convincingly replace the criterion of "intermittently empirical problem shifts". Programmes using these criteria for theory assessment cannot be considered progressive although the underlying theories seem fruitful and promising. Restoring progress in the Lakatosian sense is thus impossible, although the assignment of degeneration does not seem adequate (if so, why develop new methods to test these theories?). This is precisely where the third category enters the stage.    \\

We emphasize again: the diagnosis of stagnation preserves a research programme from being prematurely discarded, aligning with the Lakatosian perspective. Stagnation is not simply a sort of slowed empirical progression, revealing some positive signals, but neither a delayed degeneration: stagnant programmes have be revitalized first to show at least some dynamics again (in both a promising or discouraging way, allowing for their evaluation again). With the four introduced criteria for stagnation fulfilled, one confirms to active researchers a sound and inherently promising methodology, hindered only by certain external constraints. Diagnosing stagnation provides an explanation for the absence of empirically progressive shifts without diminishing the scientific value of the underlying or attributing degeneration to it. It thus completes the spectrum of evaluation, which, without this third option, might overlook or misattribute a significant part of today's scientific reality, thereby unjustifiably degrading good scientific practice. Stagnation, within the Lakatosian framework, offers a theoretical justification for maintaining topics of paramount importance for a discipline and occasionally providing additional support, as the time for an eventual breakthrough, though in the distant future, no longer seems implausible. The third option ensures that the research programme has been thoroughly examined for potentially degenerative elements and is grounded in good scientific practice. Moreover, the discussed problem of non-decidability (progress or degeneration?) in the original work is addressed. Another advantage of the introduced trinity is that the Lakatosian model becomes considerably more applicable to current issues. By conceding that stagnant phases may be lengthy but not endless, the evaluation is not predominantly left to future historians but allows for a more nuanced assessment of the status quo. 

\subsection{Outlook: The Threat of Exhaustive Stagnation in Future Research}
 An important question remains for the final justification of adding a third state of research programmes to the Lakatosian binary distinction of progress and degeneration: what signs, beyond individual examples\footnote{Several other research programmes can be identified go beyond the strict framework of stagnant programmes: the transition from non-living to living matter remains an unresolved puzzle in biology to this day. The cure for certain diseases such as Parkinson's or diabetes has also been a decades-long endeavor with limited success. Successes in non-curative therapy over recent decades must, of course, be clearly distinguished from ultimate cures. Lastly, the mystery of consciousness and the emerging qualia should be mentioned, one of the oldest problems in philosophy and later in natural science, which remains almost entirely unanswered to this day. While neuroscience has provided many functional explanations, the actual problem remains untouched: "Consciousness, however, is as perplexing as it ever was. (...) We do not just lack a detailed theory, we are entirely in the dark about how consciousness fits into the natural order," D.\;Chalmers (1996) writes.}, indicate the fruitfulness of introducing stagnant research programmes as an archetypal state of scientific practice?  
The purpose of this extended outlook is to explore the concept of stagnation beyond its strict application to individual programmes with foundational physics as a prime example, towards a more comprehensive view of fading progress across the sciences. By generously extending the concept of stagnation, the outlook invites future exploration of how the principles of the extended MSRP might adapt to address these large-scale challenges, even if the individual cases listed may not fully meet the narrower four conditions of stagnation. This extension is meant to provoke further discussion and reflection on the role of stagnation in contemporary scientific practice that shall be dealt with in greater detail in a subsequent article. This broader interpretation encompasses phenomena such as the gradual deceleration of breakthroughs, the absence of disruptive innovations, and the simultaneous experimental and theoretical stagnation across various disciplines.  While these cases may not fit the classical criteria for stagnation, they highlight trends that call for reflection on how the scientific community sustains progress amidst increasing complexity and diminishing low-hanging fruits.
For example, Jones (2009) speaks of a "burden of knowledge," somewhat corresponding to Newton's saying "standing on the shoulders of giants." The ever-longer learning process required to reach the status quo shortens productive phases, limits expertise, and brings about a dependence on collaborative work, which is not always conducive to innovation. Stagnation eventually results from the absence of innovations. 
According to Chu and Evans (2021), stagnation can arise as follows: political measures of recent decades attempted to promote scientific progress by increasing the number of publications. However, this has the opposite effect: the flood of new works can deprive reviewers and readers of the mental space required for fully grasping innovative concepts. In other words, it becomes increasingly challenging to pick out promising ideas from the cacophony of new ones and concentrate on pursuing them: "(...) too many papers published each year in a field can lead to stagnation rather than advance" (2021, p.\,1). Additionally, self-reinforcing mechanisms within research programmes regarding citation and dissemination contribute to stagnation. Stagnation is thus caused, in both cases, by a lack of attention to promising proposals. Progress comes to a halt as too much focus is placed on mostly supported theories, which may potentially be dead ends. This phenomenon can indeed be considered a variant of Kuhn's conservatism, which unnecessarily prolongs the lifespan of unproductive theories. However, stagnation should not be understood as slowing progress (unless it stagnates on such timescales that it effectively equals complete stagnation). 

Stagnation must not be understood as a slowing-down progress (unless it persists on timescales beyond Criterion 1, making it de facto equivalent to complete stagnation): a steadily decreasing general productivity, as highlighted by Bloom et al.\,(2020)\footnote{The definition of knowledge growth is based here on the simple multiplication of research productivity and the number of researchers according to Solow (1957).}, is too general an approach. Stagnation affects individual research programmes struggling with fundamental obstacles, not the symptoms of allegedly flagging research activity in general.   Let's now focus on another study supporting the concept.

Park, Leahey, and Funk (2023) analyzed 45 million publications from six decades based on the following principle: disruptive articles bring knowledge gains that influence other (sub)disciplines, leading them to be cited outside their own field. In contrast, consolidating articles with a limited sphere of influence confirm or add details to what is already known. The result: the introduced measure of disruptiveness collapses by 90 percent during the analyzed period. How does this justify the introduction of stagnant research programmes as a regularly occurring state in various disciplines? 

Disruptive articles have the potential to replace research programmes (and ignite a competition, violating Criterion 2). In the evolutionary competition described by Lakatos, the weaker, less adapted candidate is consumed by the new research programme, whose hard core has emerged from the new disruptive principles. Degeneration and elimination occur. Consolidating articles cause this competition to falter\footnote{There are specific measures to prevent this. The HRHR Policy (High-risk-high-reward) of the OECD (2023) is intended to promote the emergence of disruptive works.}. The constant affirmation of the established no longer represents progressive problem shifts, nor is the hard core at risk; instead, it becomes more entrenched. This phenomenon is similar to the self-reinforcing mechanisms highlighted by Chu and Evans (2021). If the proportion of consolidating articles becomes predominant, more and more research programmes hover between progress and degeneration. Stagnation becomes, at worst, a widespread phenomenon\footnote{It should be noted that discussions about widespread stagnation in various disciplines remain highly speculative. The inclusion of artificial intelligence could trigger a scientific revolution in a few years, advancing the progress of knowledge generation and processing. The OECD lists numerous application fields and opportunities for artificial intelligence in various research programmes in (Nolan 2021). Nevertheless, it is unclear whether it can offer a way out of stagnation, especially regarding the fundamental intellectual or empirical hurdles that can arise in stagnant programmes. 
First successes in mathematical and logical thinking, essential for the development of new theories, such as foundational physics, are shown in (Trinh et al. 2024).}. However, it must be emphasized that the results presented in (Park, Leahey, Funk 2023) at most give indications to diagnose universal stagnation.

Regarding the analytically derived criteria, the following additional systemic criteria can be added especially if one is willing to apply the concept of stagnation beyond foundational physics:
\begin{itemize}
\item  Regarding Criterion 2 of singularity: not only isolation due to lacking alternatives, but also the opposite scenario may occur: a flood of competing works that cannot be adequately grasped simultaneously. The consequence is equal to a lack of alternatives - a coexistence of many isolated, unevaluated programmes.
\item  Regarding Criterion 3 of benign epistemic limits: the research programme is already highly developed and problems therein cannot be resolved by single minds, requiring collaborative teamwork that may inhibit innovation.
\item  Regarding Criterion 4 of scientific value: the initially promising research programme receives little attention compared to others for sociological reasons (unjustified insignificance).
\end{itemize}

\section{Conclusion}
With the introduction of a third, neutral (and thus unevaluable) state of stagnant research programmes, we aimed to present a mutually exclusive and collectively exhaustive categorization scheme, which now maps the entire landscape of research programmes, thereby joining four (two current) discussions in philosophy of science. 
\begin{enumerate}
\item[i)] The debate about post-empirical science, initiated by Dawid (2014): post-empiricism might be even, once elaborated and accepted, a suitable answer how to treat programmes doomed to (eternal) stagnation without revealing signs for degeneration. But even if empiricism turns out the be the only way to confirm theories once and for all, our concept illustrates that the evaluation of "Big Science" projects is complex and cannot only distinguish between success and failure (presupposing ongoing dynamics in every research programme).
\item[ii)] The debate about fundamental physics in crisis, initiated in the broader public by, among others, Lindley (1993) and Hossenfelder (2018): the option of stagnant research programme is a moderate opposition to the sharp criticism from which a degeneration of the entire branch of fundamental physics must be deduced (given the numerous theories extraordinarily hard to test). While several aspects of their critique about too speculative theories is justified, the presented criteria for stagnation offer a more nuanced perspective on today's research landscape in that realm, especially taking benign epistemic limits into consideration: stagnant programmes as a linguistic empowerment for the majority of researchers that justifiably denies empirically progressive states in BSM physics or quantum gravity.
\item[iii)] The debate about definitional vagueness in the Lakatosian framework, initiated by Feyerabend (1975): as a side effect, progressive, stagnant and degenerative programmes finally become mutually exclusive: in an attempt to rescue the classical, dichotomous categorization by specifying a concrete time frame, the absence of progress within this time frame would automatically lead to the diagnosis of programme degeneration without further justification. The historical definitional generosity by Lakatos himself seems inevitable in this case. If, however, degeneration is not the only option missing experimental success can lead to, but a neutral stagnation as well, an unambiguous classification of research programmes should always be possible.
\item[iv)] The incorporation of stagnation brings us closer to Lakatos' original normative intentions, restoring the framework’s role as a guide for evaluating the vitality of research programmes (allowing for more than ex-post evaluations of programmes and bringing the judgment into the present again). In today’s complex and resource-intensive scientific landscape, the need for such normative guidance is more crucial than ever. The extended MSRP offers a framework for making these judgments, ensuring that promising research is not prematurely discarded while acknowledging the realities of stagnation in the scientific environment of the 21st century.
\end{enumerate} 
Despite formal motivation for introducing stagnation as the third state of a research programme, the main motivation ultimately comes from practice showing that the new categorization goes beyond an academic gimmick, but addresses real issues in philosophy of science.
In the end, there are good reasons why thousands of physicists are still spending their lifetimes on research programmes which, according to Lakatos, can in no way be considered progressive, and in which the occasional empirical problem shift may be forever absent. Can they all be wrong, among them possibly the brightest minds in their discipline? They are doing the opposite of representatives of a degenerative programme. They carry on. They dedicate themselves to stagnant programmes.
\newpage
\noindent \textbf{Bibliography} \\ \\ 
\begin{small}
\begin{onehalfspace}
Bloom, N., Jones, C. I., Van Reenen, J., Webb, M. (2020): Are ideas getting harder to find?  \textit{Am.Econ. Rev.} 110, 1104–1144.  \url{https://doi.org/10.1257/aer.20180338}\\ 
Bokulich, A.: \textit{Open or closed? Dirac, Heisenberg, and the relation between classical and quantum mechanics.} Studies in History and Philosophy of Science Part B 35(3): 377-396, 2004 \\
Bury, J. B. (1932): \textit{The Idea of Progress: An Inquiry into Its Origin and Growth.} Access via Project Gutenberg\\  
Chalmers, D. (1996): \textit{The Conscious Mind: In Search of a Fundamental Theory.} New York: Oxford University Press \\ 
Chu, J.S.G., Evans, J.A. (2021): Slowed canonical progress in large fields of science. \textit{Proc. Natl. Acad. Sci.} USA 118. \url{https://doi.org/10.1073/pnas.2021636118} \\ 
Craig, N. (2022): Naturalness. A Snowmass White Paper. \url{https://doi.org/10.48550/arXiv.2205.05708}. Submitted to the \textit{Proceedings of the US Community Study
on the Future of Particle Physics} \\
Dawid, R. (2013): \textit{String Theory and the Scientific Method.} Cambridge University Press \\ 
Dawid, R. (2019): The Significance of Non-Empirical Confirmation in Fundamental Physics. In: Dardashti, R., Dawid, R., Thebault, K. (eds.): \textit{Why Trust a Theory? Epistemology of Modern Physics.} Cambridge University Press, 99-119 \\ 
Feyerabend, P. (1975): \textit{Against Method.} London: Verso Books\\ 
Funk, R., Leahey, E., Park, M. (2023): Papers and patents are becoming less disruptive over time. \url{https://doi.org/10.1038/s41586-022-05543-x}. \textit{Nature} 613: 138–144 \\ 
Giudice, G.F. (2017): The Dawn of the Post-Naturalness Era. \url{https://doi.org/10.48550/arXiv.1710.07663} Contribution to the volume \textit{From My Vast Repertoire - The Legacy of Guido Altarelli}. arXiv:1710.07663  \\  
Harlander, R., Martinez, J., Schiemann, G. (2023): The end of the particle era?  \textit{Eur.\,Phys. H} 48(6). \url{https://doi.org/10.1140/epjh/s13129-023-00053-4}\\  
Heisenberg, W.: \textit{The Correctness-Criteria for Closed Theories in Physics.}
In: \textit{W. Heisenberg Encounters with Einstein: And Other Essays on People, Places, and
Particles.} Princeton, NJ: Princeton University Press, 1972 \\
Horgan, J.: The End of Science (1996): Facing The Limits Of Knowledge In The Twilight Of The Scientific Age. Basic Books \\
Hossenfelder, S. (2018): \textit{Lost in Math: How Beauty Leads Physics Astray.} S. Fischer Verlag\\ 
Hossenfelder, S. (2019): Good Problems in the Foundations of Physics, Blog entry in backreaction.blogspot.com \\ 
Hossenfelder, S. (2021): Screams for Explanation: Finetuning and Naturalness in the Foundations of Physics.\textit{Synthese} 198: 3727–3745. \url{https://doi.org/10.1007/s11229-019-02377-5}  \\ 
Huggett, N. (2014): Review of "String theory and the scientific method" by Richard Dawid. \textit{Notre
Dame Philosophical Reviews} \\
Jones, B. F. (2009): The burden of knowledge and the ‘death of the renaissance man’: is
innovation getting harder? \textit{Rev. Econ. Stud.} 76: 283–317 \\ 
Kuhn, T.: \textit{Objectivity, value judgment, and theory choice}.
In: David Zaret (ed.) \textit{Review of Thomas S. Kuhn The Essential Tension: Selected Studies in Scientific Tradition and Change.} Duke University Press. 320-339, 1981 \\
Lakatos, I.: Falsification and the Methodology of Scientific research programmes. In: Lakatos, I., Musgrave, A. (1970): \textit{Criticism and the Growth of Knowledge.} Cambridge University Press \\
 Lakatos I. (1976): History of science and its rational reconstructions. In: Howson, C. (ed.): \textit{Method and Appraisal in the Physical Sciences: The Critical Background to Modern Science}, 1800–1905. Cambridge University Press\\ 
Lindley, D. (1993): \textit{Das Ende der Physik. Vom Mythos der Großen Vereinheitlichten Theorie.} Springer, Basel \\ 
Massimi, M. (2021): Investing in fundamental research: for whom? A philosopher’s perspective. In: Beck, H., Charitos, P.: \textit{The Economics of Big Science. Essays by leading scientists and policy makers}, Science Policy Reports Series (Springer), 113–116. \url{https://doi.org/10.1007/978-3-030-52391-6_16} \\ 
Nolan, A. (2021): Artificial intelligence and the future of science. OECD 2021 \\ 
OECD (2023): Effective Policies to Foster High-risk/High-reward Research. OECD Science, Technology,
and Industry Policy Papers \\ 
 Oriti, D. (2009): \textit{Approaches to Quantum Gravity – Toward a New Understanding of Space, Time and Matter.} Cambridge University Press, Cambridge \\ 
Rescher, N. (1978): \textit{Scientific Progress: A Philosophical Essay on the Economics of Research in Natural Science.} University of Pittsburgh Press  \\ 
Rovelli, C.; Smolin, L. (1987): Knot Theory and Quantum Gravity. \url{https://doi.org/10.1103/PhysRevLett.61.1155}.\textit{ Phys.\,Rev.\,Lett.}\,61(10): 1155–1158 \\  
Shiltsev, V. D. (2020): Ultimate Colliders for Particle Physics: Limits and Possibilities. In: Chattopadhyay, S. et al.
(eds.) \textit{Beam Acceleration in Crystals and Nanostructures.}  Singapore: \textit{World Scientific}, 5–12. \url{https://doi.org/10.1142/9789811217135_0002} \\
Solow, R. (1957): Technical Change and the Aggregate Production Function. \textit{Review of
Economics and Statistics} 39(3): 312–320 \\ 
Stent, G. (1969): \textit{The Coming of the Golden Age: A View of the End of Progress.} Garden City, New York\\ 
Stent, G. (1978): \textit{Paradoxes of Progress.} San Francisco (Freeman, W.) \\ 
Trinh, T. et al. (2024): Solving olympiad geometry without human demonstrations. \textit{Nature} 625: 476.\url{https://doi.org/10.1038/s41586-023-06747-5}  
\end{onehalfspace}
 \end{small}





\end{document}